%% file: master.tex
\documentclass[conference,a4paper]{IEEEtran}
\IEEEoverridecommandlockouts

\usepackage{cite}
\usepackage{amsmath,amssymb,amsfonts}
\usepackage{graphicx}
\usepackage{textcomp}
\usepackage{xcolor}

\usepackage{tikz}
\usepackage{pgfplots}
\usetikzlibrary{calc}

\usepackage{cleveref}
\usepackage{csquotes}
\usepackage{booktabs}

\usepackage{flushend}

\definecolor{tol1}{HTML}{33bbee}
\definecolor{tol2}{HTML}{009988}
\definecolor{tol3}{HTML}{ee7733}
\definecolor{tol4}{HTML}{cc3311}

\usepackage{acro}
\DeclareAcronym{cots}{
  short = COTS,
  long = off-the-shelf,
}

\DeclareAcronym{tiff}{
  short = TIFF,
  long = tag image file format,
}

\DeclareAcronym{mse}{
  short = MSE,
  long = mean squared error,
}

\DeclareAcronym{awgn}{
  short = AWGN,
  long = additive white Gaussian noise,
}

\DeclareAcronym{leo}{
  short = LEO,
  long = low Earth orbit,
}

\DeclareAcronym{ldpc}{
  short = LDPC,
  long = low-density parity-check,
}

\DeclareAcronym{jscc}{
  short = JSCC,
  long = joint source and channel coding,
}

\DeclareAcronym{djscc}{
  short = DJSCC,
  long = deep joint source and channel coding,
}

\DeclareAcronym{snr}{
  short = SNR,
  long = signal-to-noise ratio,
}

\DeclareAcronym{cnn}{
  short = CNN,
  long = convolutional neural network,
}

\DeclareAcronym{psnr}{
  short = PSNR,
  long = peak signal-to-noise ratio,
}

\DeclareAcronym{prelu}{
  short = PReLU,
  long = parameterized rectified linear unit,
}

\DeclareAcronym{jscc-sat}{
  short = \textsc{JSCC-Sat},
  long = {joint source-and-channel coding for small satellite applications}
}

\usepackage{xspace}
\newcommand\cubesat{CubeSat\xspace}
\newcommand\cubesats{\cubesat{}s\xspace}

\newcommand\jpegtwok{JPEG\,2000\xspace}

\newcommand\sentinelii{Sentinel-2\xspace}

\begin{document}

\title{Joint Source-and-Channel Coding for Small Satellite Applications}

\author{\emph{Anonymous authors}}
\author{\IEEEauthorblockN{Olga Kondrateva}
\IEEEauthorblockA{\textit{Humboldt-Universit\"at zu Berlin}\\
Berlin, Germany \\
kondrate@informatik.hu-berlin.de}
\and
\IEEEauthorblockN{Stefan Dietzel}
\IEEEauthorblockA{\textit{Merantix Momentum GmbH}\\
Berlin, Germany \\
stefan.dietzel@merantix.com}
\and
\IEEEauthorblockN{Bj\"orn Scheuermann}
\IEEEauthorblockA{\textit{Technical University of Darmstadt}\\
Darmstadt, Germany \\
scheuermann@kom.tu-darmstadt.de}
}

\maketitle

\begin{abstract}
Small satellites are widely used today as cost effective means to perform Earth observation and other tasks that generate large amounts of high-dimensional data, such as multi-spectral imagery.
These satellites typically operate in low earth orbit, which poses significant challenges for data transmission due to short contact times with ground stations, low bandwidth, and high packet loss probabilities.
In this paper, we introduce \acs{jscc-sat}, which applies joint source-and-channel coding using neural networks to provide efficient and robust transmission of compressed image data for satellite applications.
We evaluate our mechanism against traditional transmission schemes with separate source and channel coding and demonstrate that it outperforms the existing approaches when applied to Earth observation data of the \sentinelii mission. 
\end{abstract}


\begin{tikzpicture}[remember picture,overlay]
  \node[
    anchor=south west,
    text width=\textwidth,
    font=\sffamily\footnotesize
  ] at ($(current page.south west) + (0.5in,1)$) {%
  DOI: https://doi.org/10.1109/LCN58197.2023.10223379\\[1ex]
  \copyright{} 2023 IEEE. Personal use of this material is permitted. Permission from IEEE must be obtained for all other uses, in any current or future media, including reprinting/republishing this material for advertising or promotional purposes, creating new collective works, for resale or redistribution to servers or lists, or reuse of any copyrighted component of this work in other works.
  };
\end{tikzpicture}

\acresetall
\section{Introduction}
Over the last decade, 
small satellites have become an increasingly important part of the space industry.
The main advantages of small satellites are their low cost and flexibility.
In particular, the \cubesat standard \cite{cubesat2020} has become popular due to availability of commercial 
\ac{cots} components, which considerably simplify and speed up the manufacturing.
Originally, the use of \cubesats was limited to demonstration flights and education purposes.
With technological advances, however, their application has expanded 
to a variety of important areas like Earth observation~\cite{rs14030589}, communication systems~\cite{radix}, 
disaster management~\cite{barmpoutis2020}, and even to deep space missions.
For example, small satellites are used as interplanetary relay for data, and they are used for monitoring the Earth's moon and planet Mars \cite{MarCO}.

Since many space applications are inherently data-centric, 
large amounts of data need to be transmitted to Earth where they are then analyzed.
Over the years, the resolution of satellite imagery has increased considerably.
Many sensors produce multi-spectral images for a wide spectrum of wavelengths, by far exceeding visible light, which results in high-dimensional image data and consequently high communication demands.
However, this demand for communication can barely be met with existing \cubesat technologies.
\cubesats are typically put into \ac{leo} and orbit the Earth several times a day.
As a result, each satellite is in communication range of a ground station only four to five times per day for short periods of around ten minutes each.
Furthermore, \cubesats are extremely power-constrained devices, 
since their small size limits the area available for solar panels.
Finally, due to severe weather conditions and the relatively high velocity of satellites, 
communication with ground stations is subject to high packet loss \cite{nogales2018}.
To properly utilize the available contact periods, 
and to compensate for satellite movement and changing weather conditions, 
a flexible and reconfigurable communication system is required.

Two major aspects of the communication system are source coding and channel coding.
Source coding is used to reduce the overall amount of data that needs to be transmitted, and channel coding protects the data against errors that result from the harsh communication environment.
Traditionally, source and channel coding are considered separately;
for example, \jpegtwok is used as a lossy compression scheme for source coding \cite{sentinel-2-user-handbook},
and \ac{ldpc} can be used for channel coding.

Although Shannon's separation theorem \cite{cover1991elements} states that optimal results can be achieved 
when source and channel coding are treated separately, 
the theorem assumes an infinite code block length, whereas, in practice, we are limited to finite lengths.
It is known that joint optimization of source and channel coding has the potential to achieve better practical communication performance \cite{6408177}.
While a number of joint coding approaches using optimization techniques exist \cite{Wei2004, Appadwedula1998, Jianfei2000}, they are too complex to be practically applied \cite{9838671}.
Recently, neural networks have been proposed to learn a \ac{jscc} scheme for transmitting RGB images in a terrestrial communication setting \cite{Bourtsoulatze2019}.

In this paper, we introduce \ac{jscc-sat}, which applies the idea of \acf{jscc} using neural networks to multi-spectral image data and adapts it to the challenging requirements of satellite communication.
Using a neural network with an encoder-decoder architecture, the input data can be directly mapped to channel symbols, combining source coding, channel coding, and modulation. 
This approach has a number of important benefits, which are particularly useful for satellite communication.
In contrast to general compression methods, 
a neural network can be trained on the specific data that is relevant for a use case, thereby taking the data's properties into account.
Benefits of using neural networks for source coding have already been demonstrated in literature \cite{Hu2022}.
As an additional benefit, \ac{jscc} mitigates the so-called \enquote{cliff effect,} a sudden loss of digital signal reception when signal strength decreases.
Moreover, \ac{jscc} offers better performance in case of varying \acp{snr} \cite{Bourtsoulatze2019}, and it adapts to different channel conditions \cite{yang2021_2}.
Finally, it can provide incremental image quality updates \cite{Kurka2021}, similar to \jpegtwok's layering technique, which encodes multiple image qualities within a single file.
Such incremental updates are particularly useful in the cases of partial transmissions due to short communication periods.

Our main contribution can be summarized as follows:
\begin{itemize}
  \item We adapt \acf{jscc} using \acf{cnn} for compression and transmission of image data in small satellite applications (\ac{jscc-sat}).
  \item We train a suitable example network using realistic \sentinelii Earth observation data and realistic signal-to-noise parameters for satellite communication.
  \item We show that \ac{jscc-sat} provides better quality than source coding using \jpegtwok combined with channel coding using \ac{ldpc} for most parameter combinations.
\end{itemize}

The remainder of this paper is organized as follows.
In \Cref{sec:related_work}, we discuss related work on satellite image compression and joint coding approaches in other domains.
We explain our application and communication setting in \Cref{sec:system_model} followed by a detailed description of \ac{jscc-sat} in \Cref{sec:jscc}.
To evaluate \ac{jscc-sat}, we compare it against separate source and channel coding in \Cref{sec:evaluation}, and we conclude the paper and give an outlook on future work in \Cref{sec:conclusion}.

\section{Related Work}
\label{sec:related_work}

In the following, we review existing mechanisms that perform \ac{jscc} with and without neural networks, in different application domains, as well as existing approaches that demonstrate the potential of neural networks for source coding alone.

\ac{jscc} received a lot of attention over the last years.
Various cross-layer optimization approaches have been proposed 
to jointly optimize the parameters of source coding, channel coding, and modulation~\cite{Wei2004, Appadwedula1998, Jianfei2000}.
A similar approach was also proposed specifically for deep-space applications~\cite{Bursalioglu2011}.
Their computation complexity, however, limits the applicability of these methods in practical scenarios.

To overcome this problem, deep \ac{jscc} based on encoder-decoder architecture has originally been proposed by Bourtsoulatze et al.~\cite{Bourtsoulatze2019}.
A benefit of \ac{jscc} using neural networks is that it mitigates the so-called \enquote{cliff effect,} a phenomenon that describes sudden significant decreases in image quality as the channel conditions change.
In subsequent works, the basic approach has been improved 
to allow for using non-differentiable channel models~\cite{Aoudia2019},
progressive transmission of images~\cite{Kurka2021}, 
transmission using a finite channel symbol alphabet (previous works assumed arbitrary complex values)~\cite{Tung2022},
orthogonal frequency division multiplexing~\cite{yang2021},
adaptive rate control~\cite{yang2021_2},
transmission of correlated sources~\cite{Xuan2021}, and many others.
All these works use general image databases for evaluation and do not take into account the specifics of satellite images.
Furthermore, the authors consider terrestrial communication channels rather than the specifics of satellite communications, such as the dependency of \acp{snr} on elevation angles.

Deep-learning-based methods were also proposed for source compression alone and are known to better preserve image quality at higher compression rates 
than traditional compression techniques do~\cite{Hu2022}.
This approach has been adopted for satellite imagery, as well.
Kong et al.~\cite{Kong2020} have proposed a spectral-spatial feature-partitioned extraction to process the spectral and spatial contents of images in parallel.
Alves de Oliveira et al.~\cite{AlvesdeOliveira2021} have introduced a lightweight variational auto-encoder suitable for on board satellite compression.
The authors have then combined onboard compression of satellite images with de-noising~\cite{9690871}, which otherwise would be performed at the ground stations.
Kong et al.~\cite{Kong2020_2} have described a multi-spectral image compression framework based on residual networks to deal with spectral and spatial redundancy.
In contrast to our work, these approaches focus only on source compression and do not take into account communication in form of channel coding and modulation. 

\section{System Model}
\label{sec:system_model}

In this paper, we assume a single \ac{leo} satellite that is used for Earth observation purposes and transmits acquired image data to a single ground station.
We further assume that the available communication channel is bandwidth constrained to an extent that requires data compression in order for transmissions to keep up with the amount of generated sensor data.

In the following, we will give a brief introduction to \sentinelii, a well known Earth observation mission that we take as basis for our evaluation,
and we give an overview of satellite hardware and its suitability to deploy neural networks.
Finally, we introduce the channel model we employ for the design of our transmission scheme.

\subsection{Earth observation using small satellites}
\label{sub:eo}

\begin{table}
  \caption{Overview of \sentinelii Bands}
  \label{tab:sentinel_bands}

  \centering
  \begin{tabular}{lll}
    \toprule
    Band & Measurement & resolution [m] \\
    \midrule
    Band 1 & Coastal aerosol & 60 \\
    Band 2 & Blue & 10 \\
    Band 3 & Green & 10 \\
    Band 4 & Red & 10 \\
    Band 5 & Vegetation red edge & 20 \\
    Band 6 & Vegetation red edge & 20 \\
    Band 7 & Vegetation red edge & 20 \\
    Band 8 & Near infrared & 10 \\
    Band 8A & Narrow near infrared & 20 \\
    Band 9 & Water vapour & 60 \\
    Band 10 & Short wave infrared: cirrus & 60 \\
    Band 11 & Short wave infrared & 20 \\
    Band 12 & Short wave infrared & 20 \\
    \bottomrule
  \end{tabular}
  
\end{table}

\begin{figure}
  \centering

  \null\vspace*{5pt}
  \includegraphics[width=\columnwidth]{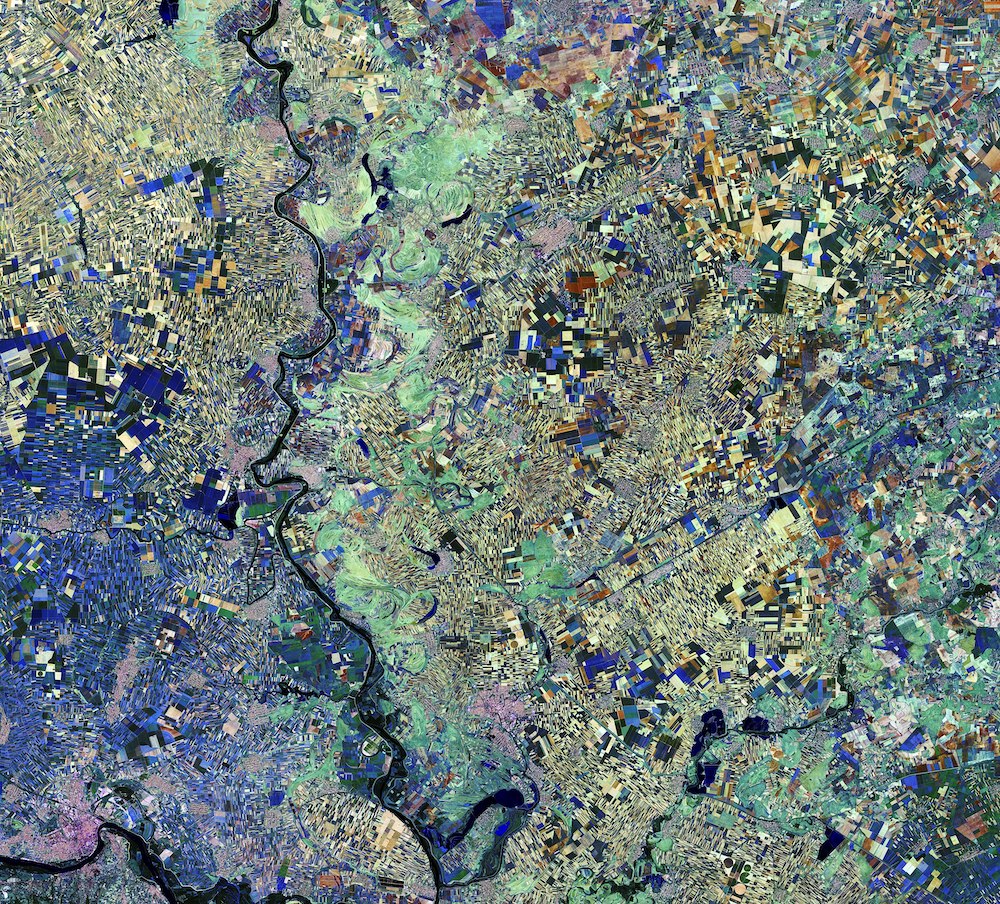}

  \caption{An example image from the \sentinelii mission showing the region of Vojvodina, Serbia (Credit: processed by ESA, CC BY-SA 3.0 IGO).}
  \label{fig:serbia}
\end{figure}

Due to their small size and relative cost efficiency, small satellites are increasingly used for earth observation tasks.
Common use cases include agricultural monitoring, emergencies management, land cover classification, and water quality estimates.
For training our neural network approach and for our evaluation, we use Earth observation data acquired from the European space agency's \sentinelii mission \cite{sentinel2}.
\sentinelii is an Earth observation mission from the Copernicus program.
Its satellites capture optical data with a relatively high resolution of $10$\,m to $60$\,m, and they operate in \acf{leo}.
Each satellite uses a multi-spectral instrument to capture not only visible light but also near infrared and short wave infrared spectra.
\Cref{tab:sentinel_bands} gives an overview of the 13 captured bands, with wavelengths ranging from approximately $430$\,nm to $2290$\,nm.
The radiometric brightness ranges up to $4096$ with a resolution of $12$\,bit.
\Cref{fig:serbia} shows an example image from the \sentinelii mission taken in 2016 \cite{vojvodina2016}, which has been recolored by analyzing the multi-spectral bands to indicate varying vegetative states, such as freshly ploughed land or different stages of crop growth or chlorophyll and water contents.

\subsection{Deploying neural networks onboard small satellites}

Since the small size of satellites limits their energy budget, it is necessary to consider energy constraints when deploying neural networks onboard of satellites.
In addressing the system's deployability and energy consumption, we turn to several key studies and missions.

The detection of cargo ships on the ocean from satellite imagery has been tested using an Nvidia TX2 SoC, which not only fits within CubeSat limitations but also sustains a total power envelope of 7.5\,W, making it compatible with small satellite missions \cite{8556744}.
The Intel Movidius Myriad 2 and STM32 Microcontroller, both low-power processors, have successfully executed star identification tasks using neural networks with a power usage of 0.89--1.08\,W for Myriad and 1.15--1.2\,W for STM32 \cite{s20216250}.
On the International Space Station (ISS), deep learning models were evaluated on the Qualcomm Snapdragon 855 and Intel Movidius Myriad X processors, further underlining the feasibility of running a neural network onboard of satellites \cite{9884906}.
A variety of models were tested that were trained on images from Earth and Mars. In particular, rather large standard ML models, such as VGG19 \cite{DBLP:journals/corr/SimonyanZ14a} and ResNet50 \cite{7780459}, were used. 
Moreover, the practical deployment of ML models onboard small satellites was tested in the $\Phi$-Sat mission via the use of Intel Movidius Myriad 2 for onboard cloud detection \cite{9600851}.

\subsection{Communication model}

\begin{figure}
  \centering
  \null\vspace*{5pt}
  \includegraphics[width=\linewidth]{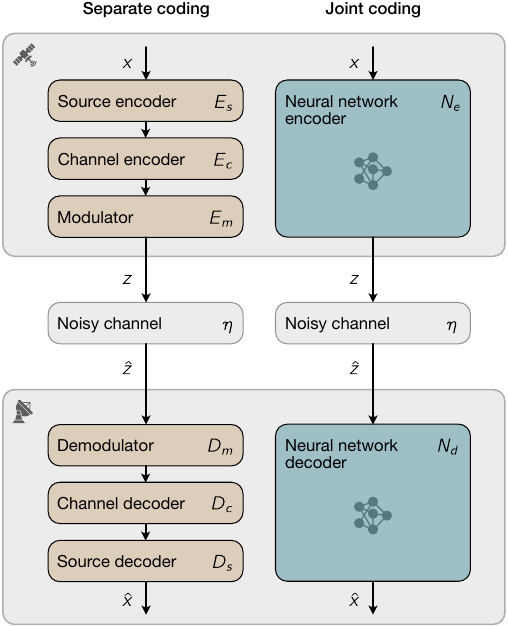}

  \caption{Comparison between the traditional communication model using separate coding and our joint coding approach.}
  \label{fig:overview}
\end{figure}

We assume a communication model where the goal is to transmit the multi-spectral images acquired by the satellite to the ground station via a bandwidth-constraint channel.
In a traditional transmission scheme, three components are considered separately, as shown in \Cref{fig:overview} on the left side.
We represent the input image $x \in \mathbb{R}^n$ as an $n$-dimensional vector.
The image $x$ is first compressed with a source encoder $E_s$, which removes the redundancies present in the input data and potentially introduces a lossy compression.
For instance, the lossy image compression mechanism \jpegtwok is often used in existing satellite communication protocols \cite{sentinel-2-user-handbook}.
Then, to ensure a reliable transmission over a noisy channel, a channel encoder $E_c$ is applied.
The channel coder adds redundancy to protect the transmission against corruption, for instance, using \ac{ldpc} codes.
Finally, during the modulation step, the output of the channel encoder is mapped to complex-valued samples $z \in \mathbb{C}^k$ using the modulation scheme $E_m$.
The resulting channel symbols are transmitted to the ground station via a noisy channel $\eta$.
Due to packet loss, the ground station receives a potentially corrupted vector of signals $\hat{z}$.
Once received, the signal is first demodulated and mapped to bits using the demodulation scheme $D_m$.
Then possibly corrupted data is restored with the help of a channel decoder $D_c$.
Finally, the data is decompressed with a source decoder $D_s$ to obtain a representation $\hat{x}$ that approximates the original image data.

For our joint coding approach (see right part of \Cref{fig:overview}), we replace the individual encoding and decoding steps with an encoder $N_e$ and a joint decoder $N_d$, which jointly optimize source (de)coding, channel (de)coding, and (de)modulation, as we will explain in the following section.

\section{Joint Source and Channel Coding for Small Satellite Applications}
\label{sec:jscc}

In order to use a neural network for joint encoding, we need to define a network architecture, which directly maps the input image $x$ to its channel representation $z$.
Our main goal is to find an encoder-decoder architecture that adapts well to a range of different channel conditions, as often observed in satellite communications.
In contrast to this joint approach, separate source and channel codes are usually chosen based on specific assumptions about the channel quality.
Therefore, their performance can degrade quickly when channel conditions differ from those expected.
Channel conditions in satellite communications, in particular, can vary quickly due to satellite movement, changing weather conditions, and wireless interference at the ground stations.
This is especially true for high-frequency communication, which is required to achieve the high data rates necessary to transmit multi-spectral images.
A similar problem occurs when the channel conditions are better than expected.
In this case, a system using separate encoders cannot improve its performance and the quality of the received data remains the same \cite{Bourtsoulatze2019}.

To design the joint system, we use an encoder-decoder neural network architecture.
Essentially, the encoder part of the network translates input image data directly to its channel representation, and the decoder translates the channel representation back to the reconstructed image data.
We train the model using multi-spectral images from the \sentinelii dataset to optimize its performance for satellite applications and use the reconstructed image's quality as reward metric during the training.
Once the training is completed, we can use the encoder part of the trained network on the satellite and the decoder part on the ground station to obtain the joint encoder and decoder, respectively.
To parametrize the network, we calculate \ac{snr} values based on suitable assumptions about the link budget for a satellite-to-ground-station link.

Next, we explain the network architecture and how we calculate suitable \ac{snr} values based on the link budget in detail.

\subsection{Neural network architecture}
\label{sec:jscc_sub}

\begin{figure*}
  \includegraphics[width=\linewidth]{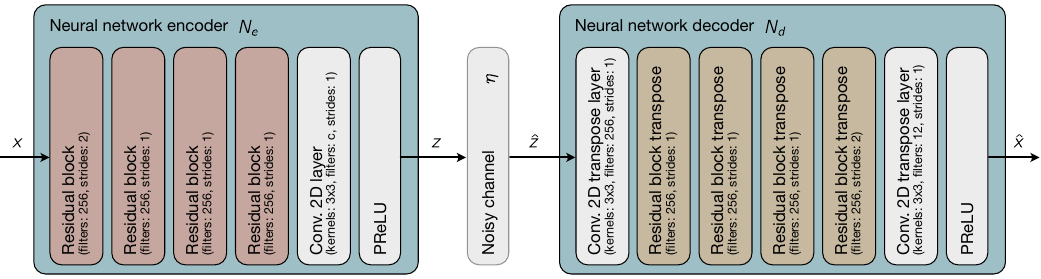}

  \caption{Encoder-decoder neural network architecture overview.}
  \label{fig:model}
\end{figure*}

\begin{figure}
  \centering
  \includegraphics[width=\linewidth]{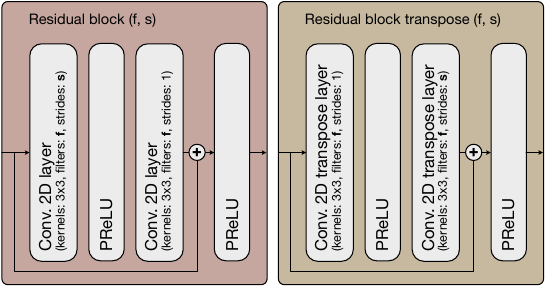}

  \caption{Residual block architectures used in the encoder and decoder parts of the neural network architecture.}
  \label{fig:residual}
\end{figure}

Our neural network architecture is based on the \ac{djscc} approach \cite{Bourtsoulatze2019}, which we adapt for the satellite communication setting.
Namely, we design a suitable neural network architecture for encoding and decoding of multi-spectral satellite images.

The system consists of a neural network serving as encoder, 
which combines source coding, channel coding, and modulation.
That is, the input image $x \in \mathbb{R}^n$ is mapped directly to channel input symbols $z \in \mathbb{C}^k$.
The ratio $k/n$ ($k < n$) determines the compression ratio of the model.
The second neural network, the decoder, receives as input the transmitted and possibly corrupted vector $\hat{z} \in \mathbb{C}^k$ 
and restores the image $\hat{x} \in \mathbb{R}^n$.

The proposed architecture is presented in \Cref{fig:model,fig:residual}.
We combine a number of layers to form both the encoder network $N_e$ and the decoder network $N_d$.
The network architecture is based on ResNet~\cite{resnet}, which is commonly used for image classification.
We adapt the network to output channel-coded symbols by replacing ResNet's dense layers, which normally perform the image classification, with a suitable layer structure for the joint encoding.
Similarly, a reversed network architecture is used for decoding.
During training, both the encoder and decoder part are trained jointly, and a non-trainable layer in between is used to model the communication channel.
During operation, the encoder and decoder parts are used separately on the satellite and ground station, respectively.
In the following, we explain the encoder and decoder network architectures in more detail.

The encoder consists of four residual blocks with 256 filters and a kernel size of $3 \times 3$.
The last layer is a convolutional layer consisting of $c$ filters, where $c$ is determined by the compression ratio $k/n$.
We use a \ac{prelu} as activation function, which generalizes the traditional ReLU by introducing a learnable parameter, which improves predictions.
The output of the encoder is combined into a vector $\tilde{z}$ of $k$ complex-valued numbers representing the channel symbols.
Then the vector is normalized as follows to ensure that the average transmit power constraint $P$ is satisfied:
\begin{equation}
  z = \sqrt{kP}\frac{\tilde{z}}{\sqrt{\tilde{z}^{*} \tilde{z}}},
\end{equation}
where $\tilde{z}^{*}$ denotes the conjugate transpose of $\tilde{z}$.
The average transmit power constraint $P$ can be intuitively understood as the maximum power that can be transmitted over the channel on average, varying depending on the channel conditions, such as distance, interference, and noise.

The next layer is a non-trainable channel layer that introduces noise to $z$ for a given \ac{snr} in order to obtain the noisy signal $\hat{z}$.
To explain the concept, we use an \ac{awgn} channel and describe how it is modelled as a layer in our encoder-decoder architecture;
we will explain how to obtain the required $\mathrm{SNR}$ values in \Cref{sub:link_budget}.
First, we compute the signal power based on the normalized channel symbols $z$:
\begin{equation}
  P_{\mathrm{sig}} = \frac{1}{k}\sum_{i=0}^{k-1}|z_i|^2
\end{equation}
Then, we determine the required noise power spectral density:
\begin{equation}
  N_0 = \frac{P_{\mathrm{sig}}}{10^{\frac{\mathrm{SNR}}{10}}}
\end{equation}
$N_0$ is used to compute the noise power:
\begin{equation}
  \sigma^2=\frac{N_0}{2}
\end{equation}
Finally, we generate a complex-valued noise vector $n$ using the normal distribution and compute the resulting noisy output:
\begin{gather}
  n = \sigma \times \bigl[\mathcal{N}(0,1) + j * \mathcal{N}(0,1) \bigr] \\
  \hat{z} = z + n
\end{gather}

Similar to encoder, the decoder consists of a convolutional transpose layer, four residual transpose blocks, another convolutional transpose layer, and a \ac{prelu} activation function.
The decoder takes $\hat{z}$ as input and restores an approximation of the image $\hat{x}$.
During training, we use the average \ac{mse} as loss function, which is defined as:
\begin{equation}
  \mathcal{L} = \mathrm{MSE} = \frac{1}{N} \sum_{i=1}^{N} d \bigl(x_i, \hat{x}_i\bigr),
\end{equation}
where $N$ is the number of samples and $d(x, \hat{x}) = || x - \hat{x} ||^2$ denotes the \ac{mse} distortion.

During evaluation, use the \ac{psnr} as a metric to determine the quality of the reconstructed image as follows:
\begin{equation}
  \mathrm{PSNR} = 10 \log_{10}\frac{\mathrm{MAX}^2}{\mathrm{MSE}},
\end{equation}
where $\mathrm{MAX}$ is the maximum possible pixel value.

Intuitively, \ac{psnr} expresses the ratio between the maximum possible signal value and the distorting noise that reduces its quality.
The metric is often used to measure the reconstruction quality of lossy compression codecs, as it approximates human perception of the reconstruction quality.

\subsection{Link budget analysis}
\label{sub:link_budget}

In order to model the transmission channel between the satellite and ground station, we need to determine the \ac{snr} values that can be assumed.
Therefore, we perform a link budget analysis in order to evaluate the communication link quality of \ac{leo} satellites with their ground stations.
In the following, we summarize the basic calculations required to compute \acp{snr}.

The expected \ac{snr} can be computed with the help of the following formula:
\begin{equation}
  \mathrm{SNR} = P_t + G_t + G_r - L - N,
\end{equation}
where $P_t$ is the transmitted power, $G_t$ and $G_r$ are the transmitter and the receiver antenna gains, respectively, $L$ is the path loss, and $N$ is the thermal noise. All quantities are calculated in decibel.
The $P_t$, $G_t$ and $G_r$ are input parameters depending on the particular equipment in use.

Finally, $L$ and $N$ need to be estimated based on the assumed environment.
$L$ can be determined from the Friis transmission formula as follows:
\begin{equation}
  L = \frac{1}{G_tG_r}\Bigl(\frac{4\pi d f}{c}\Bigr)^2,
\end{equation}
where $d$ is the slant range, that is, the distance between the satellite and the ground station, $f$ is the carrier frequency, and $c=299792458$\,m/s is the speed of light.

To compute the slant range $d$, two parameters are necessary:
the orbit hight $h$ and the elevation angle $\epsilon_{0}$.
The latter describes how high the position of the satellite is above the horizon.
Assuming the satellite flying in an overhead trajectory, the slant range can then be computed using the following formula \cite{7506756}:
\begin{equation}
  d = R_E\Biggl(\sqrt{\Bigl(\frac{h + R_E}{R_E}\Bigr)^2 - \cos^{2}\epsilon_{0}} - \sin\epsilon_{0}\Biggr),
\end{equation}
where $R_E = 6378$\,km is the radius of Earth.
The elevation angle $\epsilon_{0}$ and hence the slant range $d$ are changing over time due to the high relative movement speed of the satellite.
For our evaluation, we therefore consider a number of different elevation angles, for which we calculate \ac{snr} values and train corresponding encoder and decoder networks.
In practice, these networks can be switched based on tracking of the satellite, or they can be merged into a combined encoder and decoder network for multiple angles \cite{yang2021deep}.

Finally, the thermal noise $N$ can be computed as
\begin{equation}
  N=k \cdot T \cdot B,
\end{equation}
where $k = 1.380649 \cdot 10^{-23}$ is Boltzmann's constant, $T$ is the noise temperature, and $B$ is the bandwidth.
$T$ can be estimated as
\begin{equation}
  T = T_a + T_e,
\end{equation}
where $T_a$ is the antenna temperature, which is difficult to estimate, since it depends on weather conditions and noise level around the ground station.
We assume $T_a = 290$\,K as a suitable approximation. $T_e$ is the receiver noise temperature, that is, the noise caused by the electronics of the receiver.
$T_e$ can be estimated as
\begin{equation}
  T_e = T_0(F_{\mathrm{sys}} - 1),
\end{equation}
where $T_0=290$\,K is the reference temperature, and $F_{\mathrm{sys}}$ is the receiver's noise figure.

By using these formulae, we can approximate suitable \ac{snr} values that we plug into the encoder-decoder network's channel layer during trainings to obtain the best possible \ac{psnr} values for the given channel conditions.

\section{Evaluation}
\label{sec:evaluation}

In this section, we evaluate the \ac{jscc-sat} approach by comparing it to \jpegtwok, 
which is commonly used for compression of satellite images~\cite{sentinel-2-user-handbook}.
We use the BigEarth dataset~\cite{sumbul2019bigearthnet,Sumbul2021}, which is a collection of 590,326 multi-spectral images acquired by \sentinelii, as explained in \Cref{sub:eo}.
More specifically, we use a subset of all images contained in the dataset, which cover the area of Serbia in the summer months.
We remove cloudy images from the dataset with the help of a script provided by the BigEarthNet authors, following common evaluation practices.
In total, 14,439 multi-spectral images remain in the filtered dataset.
Each multi-spectral image is represented by 12 individual files, where each file encodes one of the bands using the lossless \ac{tiff} format.
The dataset was then further divided into training, validation, and test sets.
Since the individual bands in the dataset have different resolutions (see \Cref{tab:sentinel_bands}), we used cubic interpolation to resize all image files to the same size.
The dataset contains 12 bands. \sentinelii's band 10 contains information about clouds is not included in the dataset, since it is not useful for training.
Finally, the pixel values were normalized so that they are between 0 and 1.

To apply \ac{jscc-sat}, we implement the neural network architecture described in \Cref{sec:jscc} using Keras\footnote{Website: https://keras.io/} and Tensorfow.\footnote{Website: https://www.tensorflow.org/}
The batch size was set to 32, and the learning rate was set to $10^{-3}$ and adjusted to $10^{-4}$ after 500 epochs.
We used Adam as optimizer, which is a form of stochastic gradient descent.
A separate neural network was trained till convergence for each SNR value.

\begin{table}
  \caption{Channel Parameters}
  \label{tab:channel_parameters}

  \centering
	\begin{tabular}{ll}
		\toprule
    Parameter & Value \\
    \midrule
		Orbit height & 150\,km \\
		Carrier frequency & 2150\,MHz \\
    Transmitted power & 1\,W \\
    Satellite antenna gain & 6\,dBi \\
    Ground station antenna gain & 35\,dBi \\
		Receive channel bandwidth & 750\,kHz \\
		Noise figure & 2\,dB \\
		\bottomrule
	\end{tabular}
\end{table}

\begin{figure}
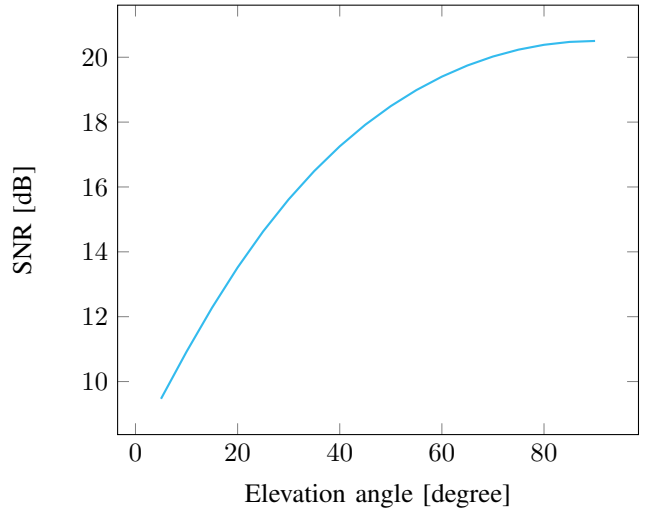

  \include{plots/plt_elevation_snr}
\vspace{-2em}
  \caption{\ac{snr} values for different elevation angles.}
\label{fig:snr_elevation}
\end{figure}

We further computed the \ac{snr} values using the formulas described in \Cref{sub:link_budget} and the parameters from \Cref{tab:channel_parameters}.
As main varying factor, we consider the satellites elevation angle relative to the ground station to cover a wide range of possible transmission environments.
\Cref{fig:snr_elevation} shows the \acp{snr} depending on the elevation angle.
A separate neural network is trained for each \ac{snr} value.

We consider two simulation scenarios:
\begin{enumerate}
  \item We compare \ac{jscc-sat} against \jpegtwok with maximum rate for source compression, that is, assuming guaranteed reliable transmission according to Shannon's separation theorem, to obtain an upper bound for the performance of \jpegtwok.
  \item We compare \ac{jscc-sat} against \jpegtwok with additional channel coding using \ac{ldpc} for a more realistic upper bound.
\end{enumerate}

For both scenarios, we compare the achieved image quality measured using \ac{psnr} values for various realistic \ac{snr} values.
The higher the \ac{psnr} values of the reconstructed images, the better the image quality.

First, we compare \ac{jscc-sat} with \jpegtwok for different compression ratios $k/n = \{ 0.33, 0.17, 0.08 \}$.
We compute an upper bound for the maximum data rate at which the data can be compressed and transmitted using Shannon's separation theorem \cite{cover1991elements}:
\begin{equation}
  R_{\max} = \frac{k}{n}C,
\end{equation}
where $C$ denotes the channel capacity, which can be computed for an \ac{awgn} channel with the help of the following formula:
\begin{equation}
  C = \log_2(1 + \mathrm{SNR}).
\end{equation}
Based on this calculation, we can obtain \ac{psnr} values for \jpegtwok assuming that no packet loss occurs, therefore constituting an upper bound.

\begin{figure}
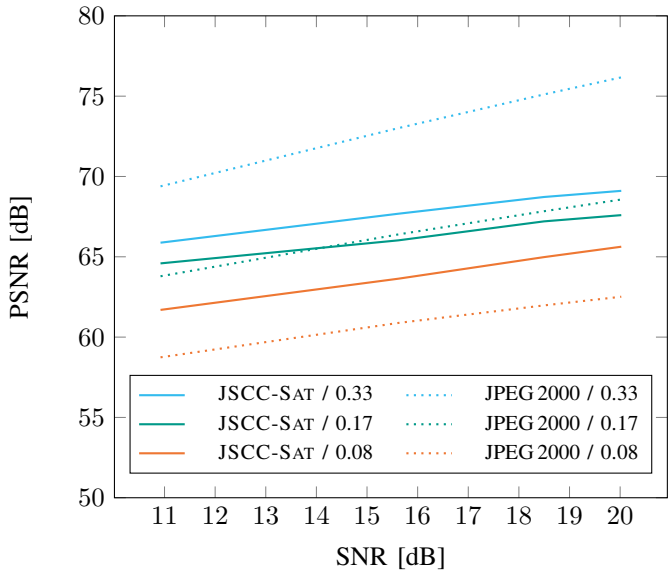

  \include{plots/plt_snr_psnr}
\vspace{-2em}
  \caption{\ac{jscc-sat} vs. \jpegtwok under optimal conditions for $k/n = \{ 0.08, 0.17, 0.33 \}$.}
\label{fig:snr_psnr}
\end{figure}

The results are presented in \Cref{fig:snr_psnr}.
The $x$-axis shows different \ac{snr} values, and the $y$-axis shows resulting \ac{psnr} values after the image data has been transmitted and reconstructed at the receiver.
It becomes clear that \ac{jscc-sat} considerably outperforms \jpegtwok for $k/n = 0.08$.
In case of $k/n = 0.17$, \ac{jscc-sat} performs better for low \ac{snr} values.
\jpegtwok, however, performs slightly better for higher \ac{snr} values with this compression ratio.
Finally, \jpegtwok performs better across all \ac{snr} values when $k/n = 0.33$.
Note again that the results for \jpegtwok describe a theoretic upper bound that is not achievable in practice.

\begin{figure}
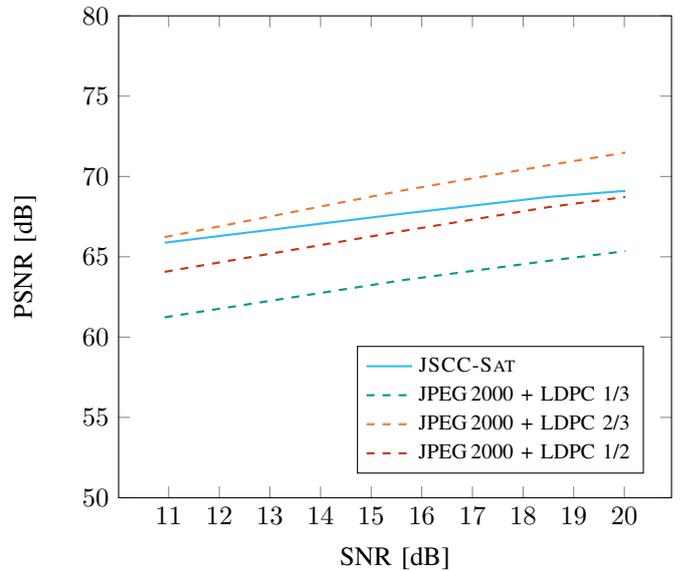

  \include{plots/plt_snr_psnr_ldpc_0.33}
\vspace{-2em}
  \caption{\ac{jscc-sat} vs. \jpegtwok + \ac{ldpc} for $k/n = 0.33$.}
\label{fig:snr_psnr_ldpc_33}
\end{figure}

\begin{figure}
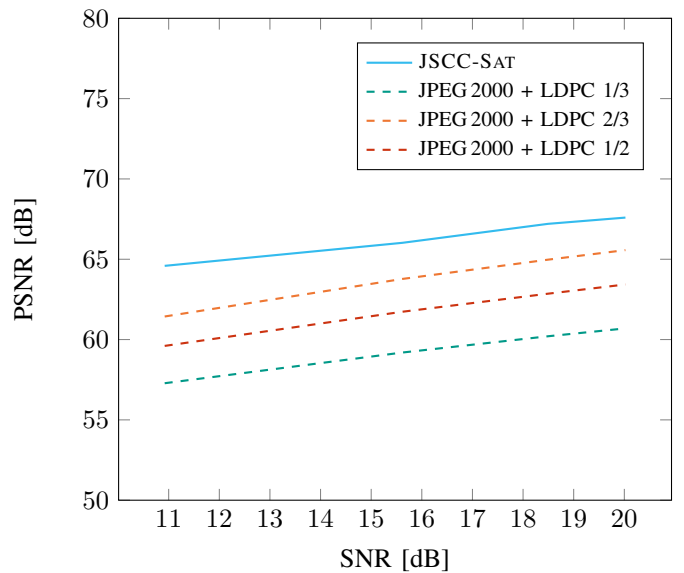

  \include{plots/plt_snr_psnr_ldpc_0.17}
\vspace{-2em}
  \caption{\ac{jscc-sat} vs. \jpegtwok + \ac{ldpc} for $k/n = 0.17$.}
  \vphantom{s}\vspace{-1pt}
\label{fig:snr_psnr_ldpc_17}
\end{figure}

\begin{figure}
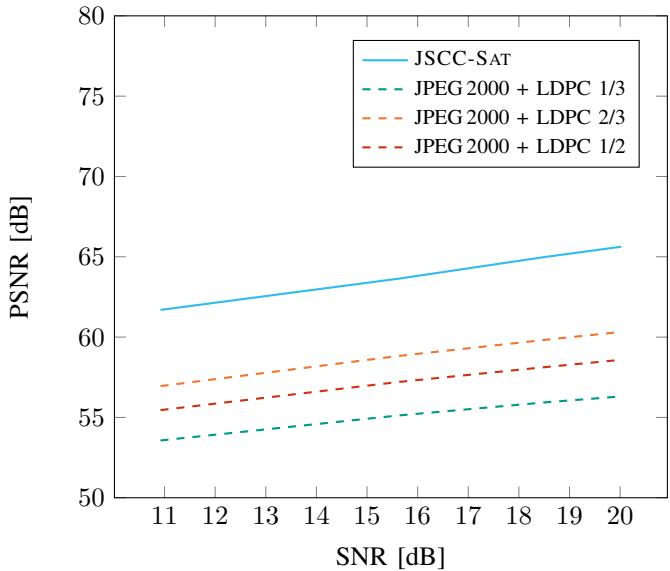

  \include{plots/plt_snr_psnr_ldpc_0.08}
\vspace{-2em}
  \caption{\ac{jscc-sat} vs. \jpegtwok + \ac{ldpc} for $k/n = 0.08$.}
\label{fig:snr_psnr_ldpc_08}
\end{figure}

To provide a more realistic upper bound, 
we apply channel coding after compressing the data with \jpegtwok.
To implement channel coding, we use the \ac{ldpc} codec with code rates $1/3$, $1/2$, and $2/3$ in our evaluation.
The results are shown in \Cref{fig:snr_psnr_ldpc_33,fig:snr_psnr_ldpc_17,fig:snr_psnr_ldpc_08}.
Each figure shows a single compression ratio $k/n$, for which we compare \ac{jscc-sat} against \jpegtwok + \ac{ldpc} with the three different code rates applied.
Like in \Cref{fig:snr_psnr}, the $x$-axis shows the \ac{snr} values and the $y$-axis the resulting reconstructed image quality measured in \ac{psnr}.

In this more realistic scenario, our \ac{jscc} approach significantly outperforms \jpegtwok + \ac{ldpc} for all code rates for compression ratios $k/n=0.08$ and $k/n=0.17$.
\Ac{jscc-sat} also performs better for code rates $1/3$ and $1/2$ for $k/n=0.33$.
The only parameter combination where \jpegtwok + \ac{ldpc} outperforms \ac{jscc-sat} is with code rate $2/3$ and compression ratio $k/n=0.33$.
This parameter combination represents the lowest compression (i.e., highest resulting transmission file size) combined with the highest channel code rate.
This combination provides the lowest data rate and thus may incur prohibitive overhead given the short visibility periods between satellites and ground stations.

Summarizing, we have shown that \ac{jscc-sat} outperforms \jpegtwok even under ideal conditions for some compression ratios, and it outperforms \jpegtwok with added \ac{ldpc} for all but the most unfavorable data rates.

\section{Conclusion}
\label{sec:conclusion}

Small satellites have recently become more and more popular due to their efficiency and support for a wide range of Earth observation applications.
In contrast to larger, more costly satellites, \ac{leo} satellites pose a number of additional communication challenges.

In particular, efficient source and channel coding schemes are required in order to cope with the large amount of generated image data that needs to be transmitted over a channel with small capacity.
While Shannon's theory states that separate mechanisms can be combined to achieve optimal results, these theoretical bounds cannot be achieved in practice.

We have considered a joint source-and-channel coding approach based on neural networks that combines source coding, channel coding, and modulation for satellite applications.
Our evaluation results show that \ac{jscc-sat} is able to outperform transmission using \jpegtwok and \ac{ldpc} for a number of realistic compression ratios covering a range of expectable \acp{snr}.
Thereby we have shown the potential to apply neural networks when designing transmission mechanisms for satellite applications to achieve more robust and flexible communication protocols.

In future work, we aim to extend our mechanism to more complex multi-satellite scenarios, where neural networks could be used to jointly optimize communication across these satellites, and we plan to investigate how to combine the trained neural networks into a single network covering a range of \acp{snr}.
Moreover, we currently focus on CNN models, which cover a wide range of common satellite applications based on vision tasks.
As an extension of our work, we see the potential to adapt our mechanism to other neural network model types that support non-vision tasks.

\bibliographystyle{IEEEtran}
\bibliography{references}

\end{document}

%% file: plots/plt_elevation_snr.tex
\begin{tikzpicture}
    \begin{axis}[
        xlabel = {Elevation angle [degree]},
        ylabel = {SNR [dB]},
        ]

        \addplot[draw=tol1,thick]
        table[x expr=\thisrowno{0}, y expr=\thisrowno{1}, col sep=space] {plots/elevation_snr};
    
    \end{axis}
\end{tikzpicture}

%% file: plots/plt_snr_psnr.tex
\begin{tikzpicture}
    \begin{axis}[
        width=\linewidth,
		height=.9\linewidth,
        ymin=50,
        ymax=80,
        xtick={11,12,13,14,15,16,17,18,19,20},
        xlabel = {SNR [dB]},
        ylabel = {PSNR [dB]},
        legend style={at={(0.5,0.05)},anchor=south,font=\footnotesize},
        legend cell align={left},
        legend columns=2,
		column sep=1em,
        ]

        \addplot[draw=tol1,thick]
        table[x expr=\thisrowno{0}, y expr=\thisrowno{1}, col sep=space] {plots/nn_comp_ratio0.33};

        \addplot[draw=tol1,thick,dotted]
        table[x expr=\thisrowno{0}, y expr=\thisrowno{1}, col sep=space] {plots/jpeg2000_comp_ratio0.33};

        \addplot[draw=tol2,thick]
        table[x expr=\thisrowno{0}, y expr=\thisrowno{1}, col sep=space] {plots/nn_comp_ratio0.17};
        
        \addplot[draw=tol2,thick,dotted]
        table[x expr=\thisrowno{0}, y expr=\thisrowno{1}, col sep=space] {plots/jpeg2000_comp_ratio0.17};

        \addplot[draw=tol3,thick]
        table[x expr=\thisrowno{0}, y expr=\thisrowno{1}, col sep=space] {plots/nn_comp_ratio0.08};

        \addplot[draw=tol3,thick,dotted]
        table[x expr=\thisrowno{0}, y expr=\thisrowno{1}, col sep=space] {plots/jpeg2000_comp_ratio0.08};
        
        \legend{
        \ac{jscc-sat} / 0.33,
        \jpegtwok{} / 0.33,
        \ac{jscc-sat} / 0.17,
        \jpegtwok{} / 0.17,
        \ac{jscc-sat} / 0.08,
        \jpegtwok{} / 0.08}
    
    \end{axis}
\end{tikzpicture}

%% file: plots/plt_snr_psnr_ldpc_0.33.tex
\begin{tikzpicture}
    \begin{axis}[
        width=\linewidth,
		height=.9\linewidth,
        ymin=50,
        ymax=80,
        xtick={11,12,13,14,15,16,17,18,19,20},
        xlabel = {SNR [dB]},
        ylabel = {PSNR [dB]},
        legend style={at={(0.95,0.05)},anchor=south east,font=\footnotesize},
        legend cell align={left},
        ]

        \addplot[draw=tol1,thick]
        table[x expr=\thisrowno{0}, y expr=\thisrowno{1}, col sep=space] {plots/nn_comp_ratio0.33};

        \addplot[draw=tol2,thick,dashed]
        table[x expr=\thisrowno{0}, y expr=\thisrowno{1}, col sep=space] {plots/awgn_ldpc_0.3333333333333333_0.33};
    
        \addplot[tol3,thick,dashed]
        table[x expr=\thisrowno{0}, y expr=\thisrowno{1}, col sep=space] {plots/awgn_ldpc_0.6666666666666666_0.33};
        
        \addplot[draw=tol4,thick,dashed]
        table[x expr=\thisrowno{0}, y expr=\thisrowno{1}, col sep=space] {plots/awgn_ldpc_0.5_0.33};
        
        \legend{
        \ac{jscc-sat},
        \jpegtwok + LDPC 1/3,
        \jpegtwok + LDPC 2/3,
        \jpegtwok + LDPC 1/2}
    
    \end{axis}
\end{tikzpicture}

%% file: plots/plt_snr_psnr_ldpc_0.17.tex
\begin{tikzpicture}
    \begin{axis}[
        width=\linewidth,
		height=.9\linewidth,
        ymin=50,
        ymax=80,
        xtick={11,12,13,14,15,16,17,18,19,20},
        xlabel = {SNR [dB]},
        ylabel = {PSNR [dB]},
        legend style={at={(0.95,0.95)},anchor=north east,font=\footnotesize},
        legend cell align={left},
        ]

        \addplot[draw=tol1,thick]
        table[x expr=\thisrowno{0}, y expr=\thisrowno{1}, col sep=space] {plots/nn_comp_ratio0.17};

        \addplot[draw=tol2,thick,dashed]
        table[x expr=\thisrowno{0}, y expr=\thisrowno{1}, col sep=space] {plots/awgn_ldpc_0.3333333333333333_0.17};
    
        \addplot[tol3,thick,dashed]
        table[x expr=\thisrowno{0}, y expr=\thisrowno{1}, col sep=space] {plots/awgn_ldpc_0.6666666666666666_0.17};
        
        \addplot[draw=tol4,thick,dashed]
        table[x expr=\thisrowno{0}, y expr=\thisrowno{1}, col sep=space] {plots/awgn_ldpc_0.5_0.17};

        \legend{
            \ac{jscc-sat},
            \jpegtwok + LDPC 1/3,
            \jpegtwok + LDPC 2/3,
            \jpegtwok + LDPC 1/2}
    
    \end{axis}
\end{tikzpicture}

%% file: plots/plt_snr_psnr_ldpc_0.08.tex
\begin{tikzpicture}
    \begin{axis}[
        width=\linewidth,
		height=.9\linewidth,
        ymin=50,
        ymax=80,
        xtick={11,12,13,14,15,16,17,18,19,20},
        xlabel = {SNR [dB]},
        ylabel = {PSNR [dB]},
        legend style={at={(0.95,0.95)},anchor=north east,font=\footnotesize},
        legend cell align={left},
        ]

        

        \addplot[draw=tol1,thick]
        table[x expr=\thisrowno{0}, y expr=\thisrowno{1}, col sep=space] {plots/nn_comp_ratio0.08};

        \addplot[draw=tol2,thick,dashed]
        table[x expr=\thisrowno{0}, y expr=\thisrowno{1}, col sep=space] {plots/awgn_ldpc_0.3333333333333333_0.08};
    
        \addplot[tol3,thick,dashed]
        table[x expr=\thisrowno{0}, y expr=\thisrowno{1}, col sep=space] {plots/awgn_ldpc_0.6666666666666666_0.08};
        
        \addplot[draw=tol4,thick,dashed]
        table[x expr=\thisrowno{0}, y expr=\thisrowno{1}, col sep=space] {plots/awgn_ldpc_0.5_0.08};
    
        \legend{
            \ac{jscc-sat},
            \jpegtwok + LDPC 1/3,
            \jpegtwok + LDPC 2/3,
            \jpegtwok + LDPC 1/2}    
    
    \end{axis}
\end{tikzpicture}